# AN IMPROVED FPGA IMPLEMENTATION OF THE MODIFIED HYBRID HIDING ENCRYPTION ALGORITHM (MHHEA) FOR DATA COMMUNICATION SECURITY

*Hala A. Farouk, Magdy Saeb*
*Arab Academy for Science, Technology & Maritime Transport*
*School of Engineering, Computer Department*
*Alexandria, Egypt*
*E-mail: mail@magdysaeb.net*

**Abstract**

*The hybrid hiding encryption algorithm, as its name implies, embraces concepts from both steganography and cryptography. In this exertion, an improved micro-architecture Field Programmable Gate Array (FPGA) implementation of this algorithm is presented. This design overcomes the observed limitations of a previously-designed micro-architecture. These observed limitations are: no exploitation of the possibility of parallel bit replacement, and the fact that the input plaintext was encrypted serially, which caused a dependency between the throughput and the nature of the used secret key. This dependency can be viewed by some as vulnerability in the security of the implemented micro-architecture. The proposed modified micro-architecture is constructed using five basic modules. These modules are; the message cache, the message alignment module, the key cache, the comparator, and at last the encryption module. In this work, we provide comprehensive simulation and implementation results. These are: the timing diagrams, the post-implementation timing and routing reports, and finally the floor plan. Moreover, a detailed comparison with other FPGA implementations is made available and discussed.*

## I. INTRODUCTION

In this work, we present an FPGA-based micro-architecture implementation of a modified version of the encryption algorithm entitled "Hybrid Hiding Encryption Algorithm (HHEA)" [SHAAR03]. In the basic version of this algorithm, no conventional substitution and translation operations on the plaintext characters are used. It rather uses simple plaintext hiding in a random bit string called the hiding vector. The name "Hybrid" is used to show that this encryption algorithm has built-in features that are inherited from data hiding techniques or "Steganography". As a matter of fact, one can use the micro-architecture for both steganography and cryptography depending on the user approach and the proper selection of the key. The basic version of this algorithm was previously implemented, as shown in reference [SAEB04a]. However, this approach did not exploit the possibility of parallel bit replacement. Furthermore, the input plaintext was encrypted serially, which caused some dependency between the throughput and the nature of the key. This dependency can be viewed by some as vulnerability in the security of the implemented micro-architecture. Based on these observations and to eliminate certain types of cipher attacks, we decided to present a modified algorithm and its accompanying micro-architecture that overcomes such limitations. The modified design eliminates the dependency between the micro-architecture throughput and the key. It also provides a significant performance improvement by fully exploiting the inherited parallelism originated by the algorithm. Moreover, the modified version escapes the chosen-plaintext attacks. In the next few sections we discuss the modified algorithm, the building blocks of the proposed improved micro-architecture along with details of its operation, the simulation and implementation results. The details of the carried out simulations, timing, routing reports and the floor plan are completely provided in the given appendix. Moreover, we present a comparison with other implementations of a selected group of encryption algorithms [SAEB02], [SAEB02].

## II. THE ALGORITHM

In the following few lines, we provide a summary of the MHHEA algorithm [SHAAR03]. The aim of the algorithm is hiding a number of bits from plain text message (M) into an N-bit long random vector (V). The locations of the hidden bits are determined by the key (K).

**Algorithm MMHHEA**
[Given a plain text message M, key matrix $K_{Lx2}$, scrambled key matrix $KN_{Lx2}$ where
$k_{ij} \in \{0,1,2,3,4,5,6,7\}$ $\begin{cases} \forall\ i = 0,\ldots, L; L \geq 15 \\ \forall\ j = 1,2 \end{cases}$
The aim of the algorithm is hiding a number of bits from plain text message (M) into a random vector (V) of bits. The locations of the hidden bits are determined by the key $K_{Lx2}$ ]
**Input:** M, $K_{Lx2}$, **Output:** encrypted file
**Algorithm Body:**
i: =0, m:=0
M[0]: =first digit in M file
**while** (M[m] ≠ EOF) [EOF: End Of File]
i: = i mod L
Generate 16-bit randomly and set them in V Vector
**if** ($K_{i,1} \geq K_{i,2}$) **then**
   z: =$K_{i,1}$
   $K_{i,1}$: =$K_{i,2}$
   $K_{i,2}$: =z
// Scramble the hiding location using the high order bits of the hiding vector
$KN_{i,1}$:=V[$K_{i,2}$+8 down to $K_{i,1}$+8] XOR $K_{i,1}$
$KN_{i,2}$= $KN_{i,1}$+($K_{i,2}$- $K_{i,1}$) mod 8
   **if** ($KN_{i,1} \geq KN_{i,2}$) **then**
      z: =$KN_{i,1}$
      $KN_{i,1}$: =$KN_{i,2}$
      $KN_{i,2}$: =z
// Scramble the message bits using the original key
q:=0
 **for** j= $KN_{i,1}$ **to** $KN_{i,2}$
    q:=q mod 3
    **if** (M[m] ≠ EOF) **then do**
        V [j]=M[m] XOR $K_{i,1}$[q]
        m: =m+1; next m in M file
    q:=q+1
  **end do**
  **next j**
Save V in output file
i: =i+1
**end while;**
**End algorithm.**

In this algorithm, we have scrambled the location and the message to overcome constant chosen-plaintext attack.

## III. THE MICRO-ARCHITECTURE

In this section we describe the micro-architecture with its operation details using a finite state machine approach (FSM). The FSM, shown in Figure 1, illustrates the conceptual required hardware modules and the elements of the design of the control unit. The machine operation takes place through six basic states. These are summarized as follows. The initial state "Init" holds back the execution of the successive states until the "Go" signal is triggered and furthermore resets all hardware modules. In the following state "LMsg", the 32-bit input plaintext is buffered for the other modules to operate on. The key is buffered into sixteen four-bit pairs of registers in the "LKey" state. The key is saved in pairs of integers. One part of the key is XOR-ed with a part of the random vector V as described in the algorithm. After the scrambling of the key, the new key points to the locations of the substitution procedure as depicted in Figure 2. In a previous work [SAEB04a], this procedure was performed serially where in each cycle one bit is replaced until the entire range from the left to right key is covered. However, we aim at designing a modified architecture that replaces the whole number of bits determined by the key in parallel rather in serial to improve the overall performance.

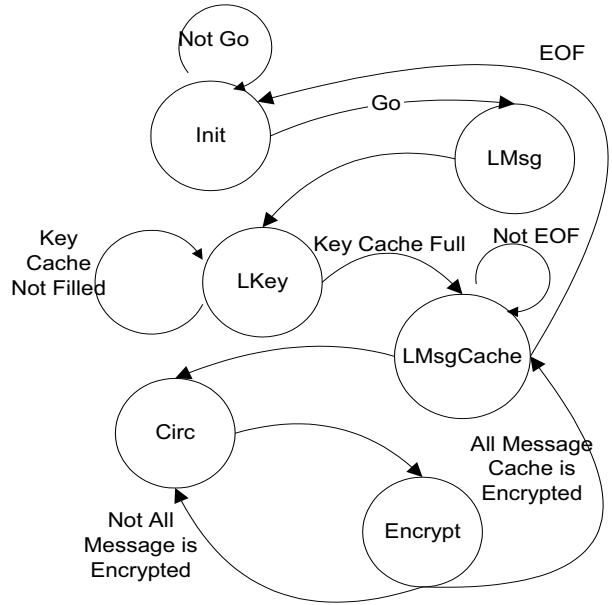

*Figure 1: The finite state machine of the micro-architecture.*

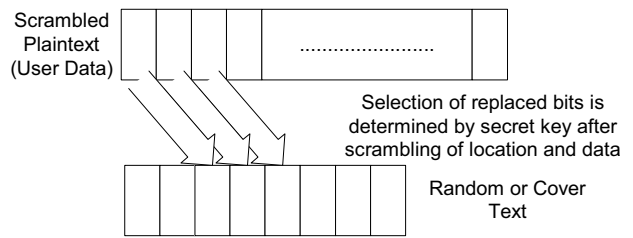

*Figure 2: The substitution procedure.*

The location of the replaced bits is determined randomly based on the generated sub-key. In this respect, two design alternatives are possible. In the first one, a variable connection between the register containing the random hiding vector and the register with the scrambled plaintext is required. Nevertheless, this approach is rather difficult. Therefore, in this modified design, the connection is fixed but the plaintext is rotated to be aligned with the bits that are to be replaced in the hiding vector. An example of the rotation scheme is illustrated in Figure 3. This leads to a considerable saving in time as well as the implementation area. The limited FPGA implementation area places a barrier on the size of the input plaintext required to be rotated. This fact has led to the splitting of the 32-bit input into two 16-bit parts. Each part is taken into a buffer inside the "Message Alignment" Module at a time during the "LMsgCache" state.

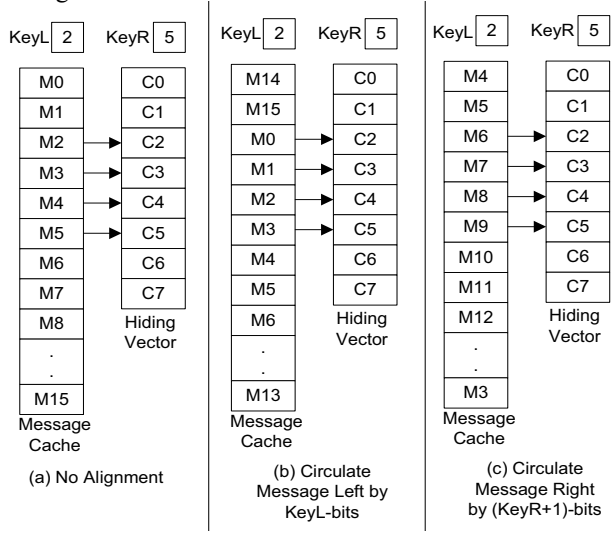

Figure 3: Message alignment using circulate left and circulate right.

The plaintext is subsequently aligned in the "Circ" state. Afterwards, the encryption or replacement procedure is performed in "Encrypt" state. These two states are interleaved in a chain of cycles until the whole 16-bit plaintext is encrypted. Consequently, the encryption process takes two clock cycles per one key pair regardless of the number of bits replaced. The micro-architecture is subdivided into six modules as shown in Figure 4. In the following subsections every module is described in details.

### 3.1 Message Cache
In this module 32-bit of the user plaintext is saved into two 16-bit registers. This is due to the fact that the "Message Alignment" module can operate on 16-bit data only. The reason for this constraint is described in the previous section.

### 3.2 Message Alignment
The "Message Alignment" module buffers the 16-bit plaintext for rotation. In order to accelerate the rotation process, multiplexers are used for n-bit rotations. Hence, the circulate operation takes only one clock cycle. In this module, the plaintext can be rotated left or right. The motivation behind this rotation procedure is explained in Figure 3. The module rotates left depending on the smaller scrambled key value and rotates right based on the larger key plus one.

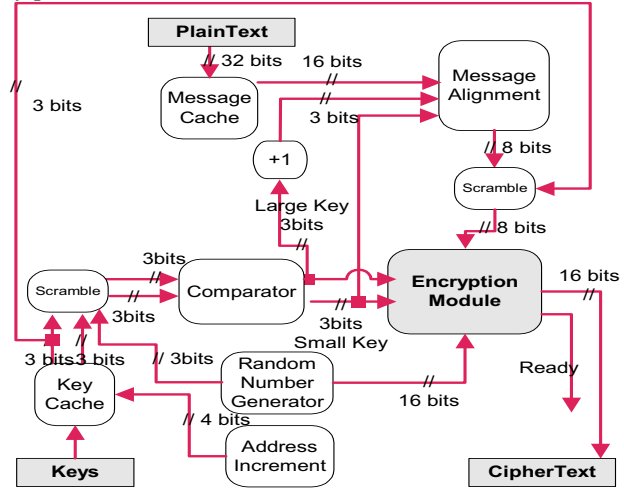

Figure 4: The block diagram of the micro-architecture.

### 3.3 Key Cache
The "Key Cache" module buffers the whole 16 three-bit key pairs. The key cache is organized as 32 three-bit registers. Each two registers share the same address to create key pairs.

### 3.4 Comparator
The comparator delivers the scrambled key with the smaller value to the "Message Alignment" module. This value is needed for the left rotation.

### 3.5 Encryption Module
The encryption module has a simple architecture of mere multiplexers that choose between the bits in the hiding vector and the ones in the scrambled plaintext stream. The selects of the multiplexers are controlled by the scrambled key pair. As a result the replacement procedure can be performed. The output cipher text is 16-bit large and is generated every two cycles. To simplify the handshaking protocol between this module and any other communication module, a ready signal is generated on every stable output.

### 3.6 Random Number Generator
The output cipher text should be scrambled as much as possible; therefore the scrambled plaintext bits are hidden inside a random string that is called hiding vector. The "Random Number Generator" module generates this hiding vector. This module is designed using Linear

Feedback Shift Register (LFSR) with primitive feedback polynomial to ensure a maximal-length sequence of random numbers.

## IV. SIMULATION

The simulation of the designed micro-architecture is performed on the Logic Simulator of the Xilinx Foundation F2.1i. The first operation performed by the MHHEA processor is loading the 32-bit plaintext during the "LMSG" state. Figure 5 depicts this operation. In this case, the input plaintext is "ABCD1234" and the "LMSG" state is active. The "LKey" state, shown in Figure 6, is the successive state to the "LMSG" state. Key pairs are loaded in parallel since they are pointed to by the same address. In the following state in Figure 7, namely the "LMSGCACHE" state, the least significant 16 bits are placed in the buffer inside the "Message Alignment" module.

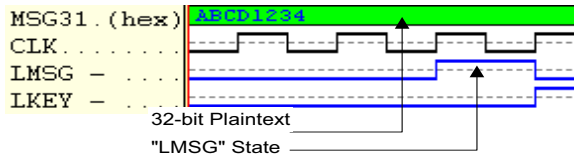

*Figure 5: Simulation of 32-bit plaintext loading.*

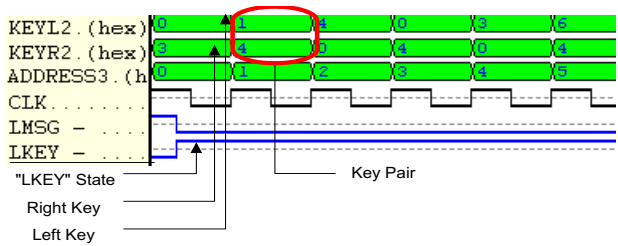

*Figure 6: Simulation of key pairs loading.*

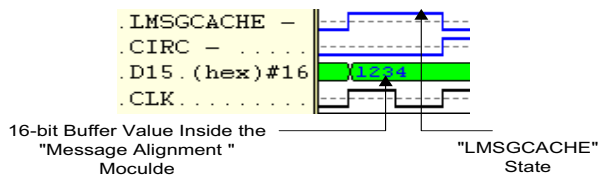

*Figure 7: Simulation of 16-bit message buffer loading.*

An example for the rotation and encryption is shown in Figure 8. The part of the key with the smaller value is, in this case, equal to zero. The other part is equal to "3". The smaller part is XOR-ed with the zero to the third location in the upper byte of the random vector V, resulting in the value "2". Adding this value to the difference between the two initial key parts, results in the second key part value, namely "5". Therefore, rotating the message twice to the left, renders the message value equal to "2341" after being "48D0". The message bits that should be encrypted are positioned from the second to the fifth location. Thus, the message is aligned with the replacement locations. Note that the location zero refers to the least significant bit. The message bits in this range of locations are equal to the hexadecimal value zero. The lower byte of the hiding vector is equal to "06". Note that in Figure 8, the lower byte of the hiding vector or the random vector is referred to as "Coverr7". The bits from the second to the fifth location in the random vector are to be replaced by the message bits in the corresponding locations to produce the cipher text equal to "CA02". The message is then rotated M times to the right, where M is equal to the larger key value plus one as mentioned before. In this case, M is equal to six. Hence, the message value "2341" is rotated to the right six times to become "048D". In this way, the least significant bits of the message buffer are always the bits yet to be encrypted.

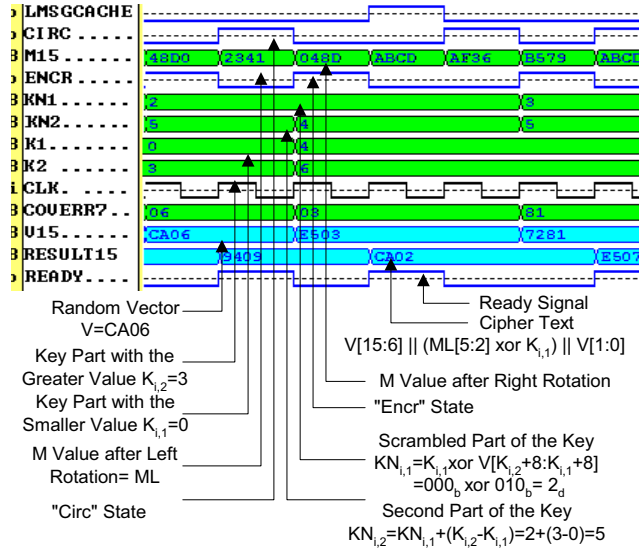

*Figure 8: Simulation of encryption process.*

## V. IMPLEMENTATION RESULTS

We have used Spartan II FPGA family to implement our design. A comparison between our micro-architecture and other encryption micro-architectures is performed through Figure 9 in Appendix A. This comparison demonstrates the dominance of the discussed algorithm and our proposed micro-architecture. This is based on the data throughput and the consumed area. The functional density is computed by the following equation:

$$\text{Funtional Density} = \frac{\text{Throughput (in Mbps)}}{\text{Area (in CLB)}},$$

The term CLB is the abbreviation for "Configurable Logic Block". In appendix B we provide a summary of the timing reports. These reports were taken for a 32-bit block plaintext and a cipher text of 16 bits. The details of these reports are as shown in Appendix A. Moreover, the floor plan of the design is also provided.

## VI. SUMMARY AND CONCLUSION

Steganography and cryptography are the essential elements of today's data communication security. Cryptography is used to scramble the data, whereas steganography is used to hide the data. The MHHEA algorithm bridges the gap between these two elements of data security. In this work, we have introduced a micro-architecture that is based on this algorithm for packet-level encryption. The special features of this micro-architecture can be summarized as follows:

- A construction that effortlessly allows the user's data block to be varied. Subsequently, the register size holding this block can be optimized depending on the implementation technology and the communication channel data rate.
- A design that allows the size of the hiding vector registers to be varied. Accordingly, a variable level of data security can be obtained. Increasing the register size leads to a higher security level. Moreover, it extends the key space with added security. The higher order byte is employed to scramble the hiding locations. Moreover, the message is scrambled using one of the key integers. This approach eliminates chosen-plain text attacks using a constant value.
- A parallel bit replacement approach that improves the overall throughput, and overcomes the security limitations encountered by a dependency between the throughput and the nature of the key.
- The micro-architecture throughput is of the order of 106 Mbps which is quite satisfactory for most of today's high speed networks.
- With a slight modification of the selected key, one can use the micro-architecture for sequential-type steganography. Moreover, if the random vector is loaded with multimedia cover data, one can immediately realize that the micro-architecture is used for hiding as well as scrambling data.
- This micro-architecture allows the user to choose between steganography and encryption by selecting the appropriate input without any changes to the hardware. Consequently, we have bridged the gap between cryptography and steganography.
- This micro-architecture can also be combined with the Steganographic Shuffler (STS), shown in [SAEB04b], for shuffled-type steganography.
- As shown in Table 1 and Figure 9, the micro-architecture provides a clear advantage when compared with other implementations. It holds the highest functional density, if we exclude the YAEA algorithm. Without a doubt, different algorithms have different degrees of security. However, we have demonstrated that with proper adaptation of the algorithm to hardware implementations, one can arrive at higher degrees of functional density and overall better performance.

The complementary nature between Cryptography and Steganography is illustrated in this work with a modified micro-architecture that can be used for both techniques. Based on the given comparison, we have demonstrated that the proposed micro-architecture shows clear performance dominance, if we exclude the variations in security levels, in data security applications of today's high speed networks.

## VII. REFERENCES

[SHAAR03] M. Shaar, M. Saeb, U. Badawi, "A Hybrid Hiding Encryption Algorithm (MHHEA) for Data Communication Security," 2003 Midwest Conference on Computers Circuits & Systems, Cairo, Egypt, 2003.
[SAEB04a] M. Saeb, M. El-Shennawy, M. Shaar, "An FPGA Implementation of the Hybrid Hiding Encryption Algorithm (MHHEA) for Data Communication Security," ICICT2004 Conference, Cairo, Egypt, 2004.
[SAEB02] M. Saeb, A. Zewail, A. Seif, "A Micro-architecture Implementation of YAEA Encryption Algorithm Utilizing VHDL and FPGA Technology," 3rd International Conference on Electrical Engineering, ICEENG, Military Technical College, Egypt, 2002.
[TRIM00] S. Trimberger, R. Pang, A. Singh, "A 12 Gbps DES Encryptor/Decryptor Core in FPGA," Lecture Notes on Computer Science, pp. 156-163, Springer-Verlag, 2000.
[GOOD00] J. Goodman, A. Chandrakasan, "An Energy - Efficient Reconfigurable Public- Key Cryptography Processor Architecture," Lecture Notes on Computer Science, pp. 175-190, Springer-Verlag, 2000.
[DAND00] Dandalis, V. K. Prasanna, J.D. P. Rolin, "A Comparative Study of Performance of AES Final Candidates Using FPGAs," A. Lecture Notes on Computer Science, pp. 125-140, Springer-Verlag, 2000.
[PATT00] C. Patterson, "A Dynamic FPGA Implementation of the Serpent Block Cipher," Lecture Notes on Computer Science, pp. 141-155, Springer-Verlag, 2000.
[SAEB04b] M. Saeb, H. Farouk, "Design and Implementation of a Secret Key Steganographic Micro-Architecture Employing FPGA," DATE2004, Designer Forum C-Lab, Paris, France, 2004.

## Acknowledgement

The authors would like to thank the reviewers for their numerous and helpful comments. The close scrutiny and constructive observations have greatly improved the final version of this paper.

## APPENDIX A

The following table [SAEB02], [SAEB02], [TRIM00], [GOOD00], [DAND00], [PATT00] and accompanying chart provide a comparison of some of the algorithms' FPGA implementations. We propose a figure-of-merit that is equal to the throughput divided by the area consumed in realizing this architecture. A chart is given below that demonstrates this figure-of-merit for some of the cited algorithms.

*Table 1: A comparison between FPGA implementations of various algorithms.*

| Algorithm | Throughput in Mbps (Taken as reciprocal of minimum period times the expected output number of information bits) | Area in CLB | Functional Density Mbps/ CLB |
|---|---|---|---|
| YAEA (XC4005xL) | 129.1 | 149 | 0.866 |
| HHEA [MARW04] | 15.8 | 144 | 0.110 |
| MHHEA [Modified] | 95.532 | 168 | 0.569 |

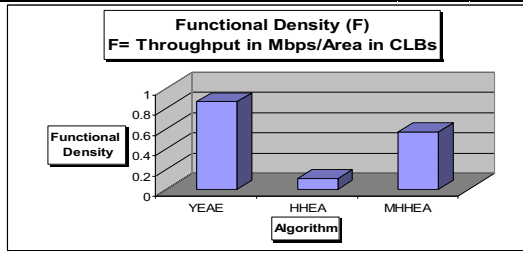

*Figure 9: The figure-of-merit of various FPGA implementations.*

### Implementation reports
In this appendix, we provide the details of the implementation reports as they were made available by the Xilinx CAD software.

**Design Information**
  Target Device                  : xc2s100
  Target Package               : tq144
  Target Speed                   : -06
  Mapper Version               : spartan2 -- C.22

**Design Summary**
  Number of Slices             : 337 out of 1200   28%
  Slice Flip Flops               : 205
  4 input LUTs                   : 393
  Number of bonded IOBs      : 57 out of 92      61%
  Number of TBUFs           : 206 out of 1280   16%
  Total equivalent gate count for design : 5051
  Additional JTAG gate count for IOBs : 2784

**Timing Summary**
  Minimum period               : 41.871ns
  Maximum frequency          : 23.883MHz
  Maximum net delay          : 6.770ns

**The Floor Plan**
The floor plan is shown in Figure 10.

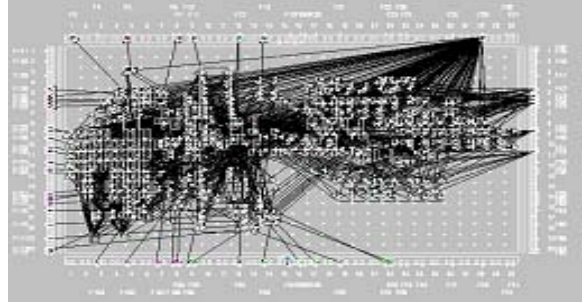

*Figure 10: The floor plan.*

## APPENDIX B

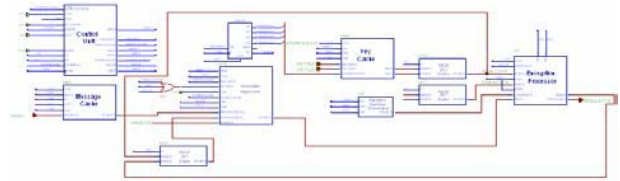

*Figure 11: The circuit diagram for the entire design.*

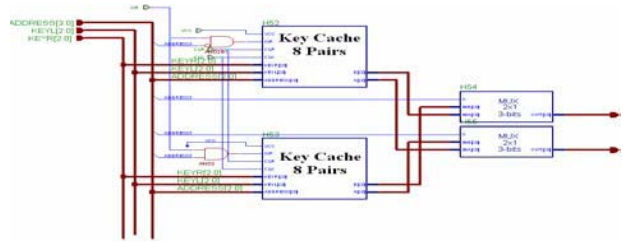

*Figure 12: The circuit diagram for the key cache.*

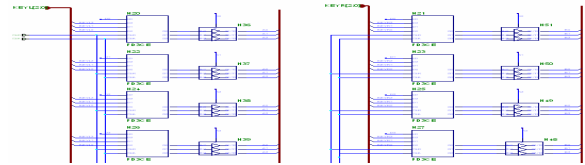

*Figure 13: The circuit diagram for the eight-pair key cache from inside.*

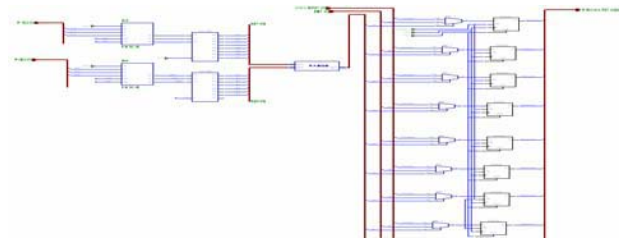

*Figure 14: The circuit diagram for the encryption module.*